\documentclass[12pt]{article}
\usepackage{epsfig}
\usepackage{amssymb,graphicx}
\def\be{\begin{equation}}
\def\ee{\end{equation}}
\def\ba{\begin{array}{c}}
\def\ea{\end{array}}
\def\p{\partial}
\def\ben{\[}
\def\een{\]}

\newcommand{\bea}{\begin{eqnarray}}
\newcommand{\eea}{\end{eqnarray}}

\newcommand{\bbr}{\br\!\br}
\newcommand{\kkt}{\kt\!\kt}

\newcommand{\pkt}{\!\!\succ\,\,}
\newcommand{\kt}{\rangle}
\newcommand{\br}{\langle}

\begin{document}

\begin{center}

{\Large \bf {

%


Quantization of Big Bang in crypto-Hermitian Heisenberg picture

 }}

\vspace{13mm}

 {\bf  Miloslav Znojil}

 \vspace{3mm}
Nuclear Physics Institute ASCR, Hlavn\'{\i} 130, 250 68 \v{R}e\v{z},
Czech Republic

 e-mail:
 {znojil@ujf.cas.cz}

\vspace{3mm}


\end{center}


%
%

\section*{Abstract}
A background-independent quantization of Universe near its Big Bang
singularity is considered. Several conceptual issues are addressed
in Heisenberg picture. (1) The observable spatial-geometry
non-covariant characteristics of an empty-space expanding Universe
are sampled by (quantized) distances $Q=Q(t)$ between space-attached
observers. (2) In $Q(t)$ one of the Kato's exceptional-point times
$t=\tau_{(EP)}$ is postulated {\em real-valued}. At such an instant
the widely accepted ``Big Bounce'' regularization of the Big Bang
singularity gets replaced by the full-fledged quantum degeneracy.
Operators $Q(\tau_{(EP)})$ acquire a non-diagonalizable Jordan-block
structure. (3) During our ``Eon'' (i.e., at all $t>\tau_{(EP)}$) the
observability status of operators $Q(t)$ is guaranteed by their
self-adjoint nature with respect to an {\em ad hoc} Hilbert-space
metric $\Theta(t) \neq I$. (4) In adiabatic approximation the
passage of the Universe through its $t=\tau_{(EP)}$ singularity is
interpreted as a quantum phase transition between the preceding and
the present Eon.

%



\section{Introduction and summary}
\label{secs1}

The recent experimental success of the measurement of the cosmic
microwave background \cite{17} resulted in an amendment of the
overall physical foundations of cosmology \cite{Mukhanov}. The
theoretical interest moved to the study of the youngest Universe
where, in the dynamical as well as kinematical regime close to Big
Bang one still has to combine classical general relativity with
quantum theory. Alas, the task looks quite formidable and seems far
from its completion at present \cite{Rovelli}.

Fortunately, even the classical, non-quantum models suffice to
describe the evolution of the Universe far from the Big Bang
singularity cca 13.8 billion years ago. It is one of purposes of our
present note to emphasize that near Big Bang, the recent progress in
quantum theory (cf., e.g., its compact review \cite{MZbook}) becomes
relevant and that it should be kept in mind with topmost
attentiveness. We believe that the impact of certain recent updates
of quantum theory upon cosmology will be nontrivial, indeed.

In what follows our main attention will be paid to the conceptual
role and increase of cosmological applicability of quantum theory
using non-standard, non-Hermitian representations of the operators
of observable quantities. Unfortunately, the terminology used in
this direction of research did not stabilize yet. In the literature
the whole innovative approach is presented under more or less
equivalent names of quasi-Hermitian quantum theory
\cite{Dyson,Geyer}, ${\cal PT}-$symmetric quantum theory
\cite{BBJ,Carl}, pseudo-Hermitian quantum theory \cite{ali} or
crypto-Hermitian quantum theory \cite{Smilga,SIGMA}.

We will discuss and analyze here the concept of quantum Big Bang in
the recently proposed crypto-Hermitian Heisenberg-picture
representation \cite{Heisenberg}. The material will be organized as
follows. Firstly, in section \ref{firstly} we shall outline the
overall cosmological framework of our considerations. Subsequently,
the basic mathematical aspects of the formalism (viz., the
crypto-Hermitian quantum theory in its three-Hilbert-space (THS)
version of Refs.~\cite{SIGMA} and \cite{timedep}) will be summarized
in sections  \ref{firstlyb} and \ref{secondly} and in an Appendix.
In section \ref{thirdly} we shall finally turn attention to several
aspects of the Heisenberg-picture quantization of our toy-model
Universe. In the last section \ref{fourthly} a few concluding
remarks will be added.

\begin{figure}                    
\begin{center}                         
\epsfig{file=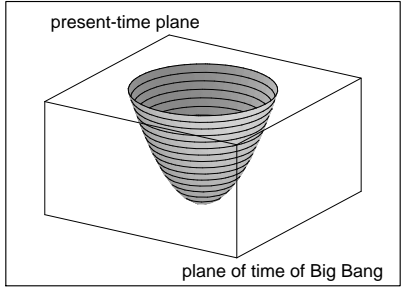,angle=270,width=0.30\textwidth}
\end{center}    
\vspace{2mm} \caption{Schematic picture of the classical expansion
of 1D Universe after Big Bang.
 \label{vizt}
 }
\end{figure}

\section{Cosmological preliminaries\label{firstly}}

There exist several imminent sources of inspiration of our present
study. The oldest one is due to Ali Mostafazadeh \cite{aliprivate}.
As early as in 2001, after my seminar talk at his University he
pointed out that the non-Hermitian but ${\cal PT}-$symmetric
Schr\"{o}dinger operators could find, via Wheeler-DeWitt equation,
an important exemplification in cosmology. Although he abandoned the
project a few years later (cf. his critical and sceptical summary of
the outcome in his review paper \cite{ali}), the idea survived. The
related necessary quantum-theoretical methods themselves are being
actively developed (cf., e.g., \cite{SIGMA} and \cite{Heisenberg}).
In what follows we intend to describe briefly both their key ideas
and their potential applicability in the Big Bang setting.



\subsection{Big Bang in classical picture}
\label{secs2s1}

\begin{figure}                    
\begin{center}                         
\epsfig{file=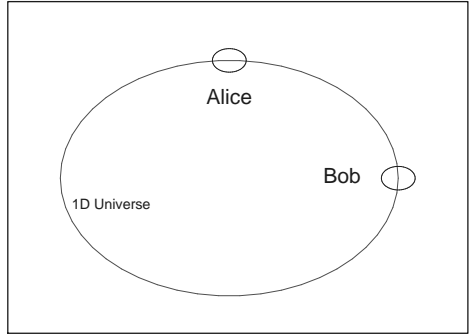,angle=270,width=0.35\textwidth}
\end{center}   
\vspace{2mm} \caption{Alice and Bob measure their 1D distance
(non-covariant idealization).
 \label{kruhym}
 }
\end{figure}

 \noindent
Our present methodical analysis of the Big Bang phenomenon cannot
have any ambition of being realistic. In a Newtonian toy model of
the evolution of an empty one-dimensional space we may visualize the
history of the Universe as a circle which blows up with time (cf.
Fig.~\ref{vizt}). The hypothetical classical observers of this
extremely simplified Universe are assumed co-moving with the space,
detecting and confirming the Hubble's law which controls the growth
of their distance $q(t)$ with time (cf. Fig.~\ref{kruhym} or pages 5
- 7 in monograph \cite{Mukhanov}).

\begin{figure}                    
\begin{center}                         
\epsfig{file=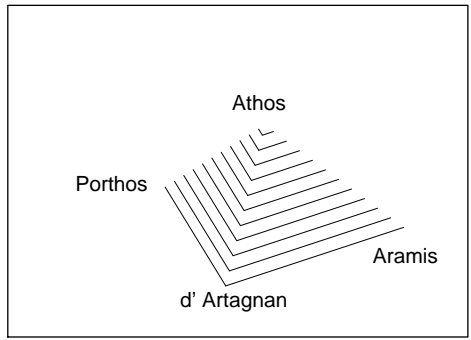,angle=270,width=0.35\textwidth}
\end{center}  
\vspace{2mm} \caption{Four non-planar musketeers measuring their
mutual distances in 3D Universe.
 \label{CTYRST}
 }
\end{figure}

After a hypothetical return to three spatial dimensions and/or to a
non-isotropic spatial geometry one will have to employ, for a
similar measurement, a non-planar quadruplet of classical observers
(cf. Fig.~\ref{CTYRST}). They may be expected to re-confirm the
Hubble's prediction of the approximate isotropy and homogeneity of
the space. Thus, for our present methodical purposes we may return
back to the 1D picture of Fig.~\ref{kruhym} and consider the
quantization of the single observable $q=q(t)$.

\subsection{The problem of survival of singularities after
quantization\label{se1}}

One of my personal most influential discussions of the quantum Big
Bang problem took place after a seminar in Paris \cite{WP} (cf. also
its published version \cite{6}) which was delivered by Wlodzimierz
Piechocki from Warsaw. In his talk the speaker analyzed the quantum
Friedmann-Robertson-Walker model in the setting of loop quantum
gravity \cite{9}. He explained why quantum theory, via Stone theorem
\cite{Stone}, seems to lead to an inevitable regularization of the
classical singularities. In the words supported by extensive
literature \cite{2}, quantization was claimed to imply the necessity
of replacement of the catastrophic dynamical Big Bang scenario by
the mere smooth process called Big Bounce.

\begin{figure}                    
\begin{center}                         
\epsfig{file=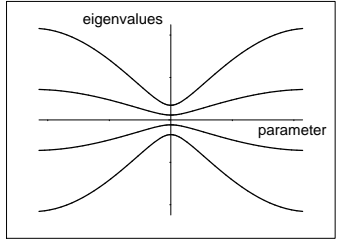,angle=270,width=0.25\textwidth}
\end{center}  
\vspace{2mm} \caption{Avoided crossing of eigenvalues; for Hermitian
matrices the phenomenon is generic.
 \label{tunebe}
 }
\end{figure}

Besides a number of physical and thermodynamical considerations
(which will not even be touched in our present text) the
mathematical essence of the latter line of argumentation is
comparatively easy to explain: In the absence of any symmetry (which
could imply an incidental degeneracy of two eigenvalues of different
symmetries) the eigenvalues of virtually any self-adjoint and
parameter-dependent operator $\Lambda(\tau)$ exhibit a ``repulsion''
as sampled in Fig.~\ref{tunebe}.

A rigorous mathematical explanation of the phenomenon is elementary:
in similar situations the coincidence of eigenvalues at a parameter
$\tau_{deg.}$ may take place if and only if this value has the
properties of the so called Kato's \cite{Kato} exceptional point,
$\tau_{deg.}=\tau_{(EP)}$. Alas, for self-adjoint operators the
value of $\tau_{(EP)}$ is necessarily complex. Thus, whenever the
parameter is time (i.e., a real variable), the evolution diagram has
{\em always} the generic avoided-Big-Bang alias Big-Bounce form of
Fig.~\ref{tunebe}.

\section{Quantum theory preliminaries\label{firstlyb}}

One of the most straightforward methods of circumventing the above
Big-Bang-avoiding paradox {\it must be sought in the use of the
time-dependent operators of observables} $\Lambda(t)$ which possess
{\em real} EP singularities. The problem becomes solvable  via a
parallel introduction of a nontrivial inner-product metric
$\Theta=\Theta(t)$ {\it which must also be necessarily
time-dependent in general} \cite{SIGMA}. Intuitively speaking, the
new degrees of freedom in $\Theta=\Theta(t)$ will suffice for an
effective suppression of the repulsive tendencies of all of the
eigenvalues of $\Lambda(t)$. In this manner, the currently accepted
hypothesis of a mathematical necessity of the disappearance of the
singularities after quantization becomes falsified.

\subsection{Quantum systems in crypto-Hermitian representation}
\label{secs3s1}

A longer version of the latter statement will form the core of our
present message. We shall demonstrate the non-universality of the
tunneling of Fig.~\ref{tunebe}. Our main task will be the transfer
of the concept of singularities from classical gravity into the
crypto-Hermitian quantum theory using the language and notation of
Ref.~\cite{SIGMA}.

\begin{figure}                    
\begin{center}                         
\epsfig{file=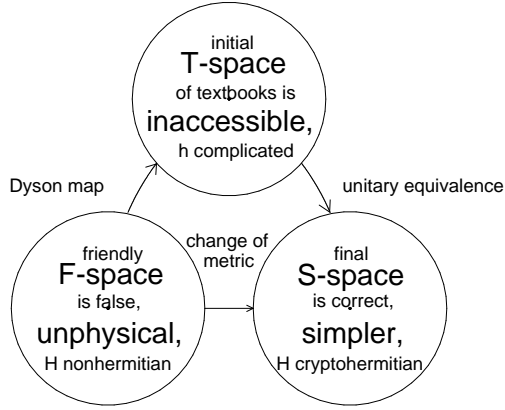,angle=270,width=0.55\textwidth}
\end{center}     
\vspace{2mm} \caption{The F-S-T triplet of representation spaces.
 \label{opo4}
 }
\end{figure}

One of the quickest introductions into such a presentation of
quantum theory using non-Hermitian representation of observables was
provided by Scholtz et al \cite{Geyer}. Within the framework of
nuclear physics these authors recalled the Dyson's \cite{Dyson} idea
that the explicit knowledge of a realistic bound-state Hamiltonian
$\mathfrak{h}=\mathfrak{h}^\dagger$ may prove useless if its
diagonalization (needed for the comparison of the theory with
experiment) proves over-complicated.

The problem and its solution emerged during the study of the
heaviest nuclei for which the self-adjoint realistic Hamiltonian
$\mathfrak{h}$ operates in a textbook Hilbert space with the
``curly-bra'' vector elements $|\psi\pkt \in {\cal H}^{(T)}$ (this
is the notation which was introduced in Table Nr. 2 of
review~\cite{SIGMA}). In the nuclear-physics literature the slow
convergence of the numerical diagonalization of $\mathfrak{h}$
proved accelerated after a {\em non-unitary} preconditioning of wave
functions,
 \be
 |\psi\pkt \ \ \to \ \ |\psi\kt = \Omega^{-1}\,|\psi\pkt \ \in
 \ {\cal H}^{(F)}\,.
 \label{precon}
 \ee
The use of an appropriate, {\it ad hoc} ``Dyson's map'' $\Omega$ and
of the friendlier ``interacting boson''
 Hilbert space ${\cal H}^{(F)}$ was recommended. The
isospectrality between self-adjoint
$\mathfrak{h}=\mathfrak{h}^\dagger$ (in ${\cal H}^{(T)}$) and its
image $H = \Omega^{-1}\mathfrak{h} \Omega \neq H^\dagger$ (which is
non-Hermitian in manifestly unphysical ${\cal H}^{(F)}$) gets
explained when one changes the inner product and when one replaces
the unphysical space ${\cal H}^{(F)}$ by its amended alternative
${\cal H}^{(S)}$.

The key features of the pattern are summarized in Fig.~\ref{opo4}.
In ``the second'' Hilbert space ${\cal H}^{(S)}$ the inner product
is constructed or chosen in such a way that the isospectral (but, in
``the first'' Hilbert space ${\cal H}^{(F)}$, non-unitary) image $H$
of the Hamiltonian $\mathfrak{h}$ (which was, by assumption,
self-adjoint in  ``the third'' Hilbert space ${\cal H}^{(T)}$)
becomes also self-adjoint. In another formulation, Hilbert spaces
${\cal H}^{(S)}$ and ${\cal H}^{(T)}$ become unitarily equivalent
and, hence, they yield the undistinguishable measurable physical
predictions.

\subsection{Stone theorem revisited}
\label{secs3s2}

In the language of mathematics the Stone theorem about unitary
evolution \cite{Stone} can be given a less common formulation even
in Schr\"{o}dinger picture in which the evolution is controlled by
Schr\"{o}dinger equation
 \be
 {\rm i}\,\partial_t|\psi\kt=H\,|\psi\kt\,
 \label{SEtd}
 \ee
(here, $H$ must have real and discrete spectrum, usually also
bounded from below). The {\em unitary\,} evolution of ket vector
$|\psi\kt$ {\em may still be reestablished} even for $H \neq
H^\dagger$ when using an {\em amended} inner-product metric $\Theta
\neq I$. A non-equivalent Hilbert space ${\cal H}^{(S)}$ of the
preceding paragraph is obtained in this manner.

The construction enables us to define a new operator adjoint
$H^\ddagger= \Theta^{-1}H^\dagger \Theta$. Under certain natural
conditions {\em the same} Hamiltonian $H$ may be then declared
self-adjoint in ${\cal H}^{(S)}$ whenever the metric is such that
$H=H^\ddagger$. Some of the necessary mathematical properties of the
Hamiltonian-Hermitizing metric operator were thoroughly discussed in
\cite{Geyer}. Their rigorous study may also be found in the recent
edited book \cite{book} and, in particular, in its last chapter
\cite{ATbook}.

The sense of the whole recipe is in rendering the evolution law
(\ref{SEtd}) {\em unitary} in ${\cal H}^{(S)}$, i.e., fully
compatible with the first principles of quantum mechanics. In other
words, a {\em unitary} evolution of a quantum state in ${\cal
H}^{(S)}$ may be {\em misinterpreted} as non-unitary when studied in
an ill-chosen Hilbert space ${\cal H}^{(F)}$ in which the
Hamiltonian is not self-adjoint,  $H \neq H^\dagger$ \cite{ali}.

\subsection{Unconventional Schr\"{o}dinger picture}
\label{secs3s3}

In the the conventional Schr\"{o}dinger picture (SP) the Hamiltonian
$\mathfrak{h}_{(SP)}(t)$ is assumed self-adjoint in a textbook-space
${\cal H}^{(T)}$. It may be assumed to generate also the unitary
evolution of the wave functions $|\psi(t)\pkt$ of the Universe.
Still, in the light of the preceding two paragraphs this generator
may prove {\em simplified} when replaced by its isospectral,
Big-Bang-passing (BBP) partner
 \be
 H_{({BBP})}(t)=\Omega^{-1}_{({BBP})}(t)\,
 \mathfrak{h}_{(SP)}(t)
   \, \Omega_{({BBP})}(t)
    \,.
   \label{haham}
 \ee
One could choose here {\it any} (i.e., in general, non-unitary and
manifestly time-dependent) invertible Dyson's operator
$\Omega_{({BBP})}(t)$ which maps the initial physical Hilbert space
${\cal H}^{(T)}$ on its (in general, unphysical, auxiliary) image
${\cal H}^{(F)}$. Subsequently, one defines the so called physical
metric
 \be
 \Theta_{(BBP)}(t)=
 \Omega_{(BBP)}^\dagger \!(t)\,\,\Omega_{(BBP)}(t)\,.
 \label{fakto}
 \ee
The desired amendment of the unphysical inner product is achieved
\cite{ali}. Indeed, it might look rather strange that we are now
dealing with a time dependent scalar product, but an exhaustive
explanation and resolution of the apparent paradox has been provided
in Ref.~\cite{SIGMA}. In a way summarized in Fig.~\ref{opo4} above
one merely returns from the auxiliary Hilbert space ${\cal H}^{(F)}$
to its ultimate physical alternative  ${\cal H}^{(S)}$. By
construction, the latter one is ``physical'', i.e., unitarily
equivalent to the initial one, ${\cal H}^{(S)} \sim {\cal H}^{(T)}$.

We are now prepared to make the next step and to return to the
problem of the cosmological applicability of the whole
representation pattern of  Fig.~\ref{opo4} as summarized briefly
also in subsection \ref{secs3s1}. First of all we have to take into
consideration the manifest time-dependence of our model-dependent
and geometry-representing preselected observable
$Q(t)=Q_{({BBP})}(t)$. This operator is defined in both ${\cal
H}^{(F)}$ and ${\cal H}^{(S)}$. Via an analogue of Eq.~(\ref{haham})
the action of this operator may be pulled back to the initial
Hilbert space ${\cal H}^{(T)}$, yielding its self-adjoint avatar
 \be
 \mathfrak{q}_{(SP)}(t)=\Omega_{({BBP})}(t)
    \,
 Q_{({BBP})}(t)\,\Omega^{-1}_{({BBP})}(t)\,.
   \label{kuham}
 \ee
In this manner, the observability of $Q_{({BBP})}(t)$ is guaranteed
if and only if
 \be
 Q^\dagger_{({BBP})}(t) \Theta_{({BBP})}(t)
 =\Theta_{({BBP})}(t)\,Q_{({BBP})}(t)\,.
 \label{thisco}
 \ee
The latter relation may be re-read as a linear operator equation for
unknown $\Theta_{({BBP})}(t)$. When solved it enables us to
reconstruct (and, subsequently, factorize) the metric which we need
in the applied BBP context.

In the next step of the recipe of Ref.~\cite{SIGMA} our knowledge of
the time-dependent operator (\ref{haham}) and of the Dyson's map
$\Omega_{({BBP})}(t)$ enables us to introduce a new operator $
G_{({BBP})}(t) = H_{({BBP})}(t) - {\Sigma}_{({BBP})}(t)$ where
 \be
 {\Sigma}_{({BBP})}(t)
 = {\rm i}\Omega^{-1}_{({BBP})}(t)\left [\partial_t\Omega_{({BBP})}(t)\right ]
 \,.
 \label{diffe}
  \ee
The SP evolution of wave functions in ${\cal H}^{(F)}$ and ${\cal
H}^{(S)}$ will then be controlled by the pair of Schr\"{o}dinger
equations of Ref.~\cite{SIGMA},
 \be
 {\rm i}\p_t|\Psi^{({BBP})}(t)\kt = G_{({BBP})}(t)\,|\Psi^{({BBP})}(t)\kt\,,
 \ \ \ \ \ \
 |\psi^{({BBP})}(t)\kt \in {\cal H}^{(F)}_{({BBP})}\,,
 \label{SEtaujoa}
 \ee
 \be
 {\rm i}\p_t|{\Psi}^{({BBP})}(t)\kkt =
  G_{({BBP})}^\dagger(t)\,|{\Psi}^{({BBP})}(t)\kkt\,,
 \ \ \ \ \ \
 |{\psi}^{({BBP})}(t)\kkt \in {\cal H}^{(F)}_{({BBP})}\,.
 \label{SEtaujobe}
 \ee
We may conclude that the time-dependence of mappings
${\Omega}_{({BBP})}(t)$ does not change the standard form of the
time-evolution of wave functions too much. One only has to keep in
mind that the role of the generator of the time-evolution of the
wave functions is transferred from the hiddenly Hermitian ``energy''
operator $H_{({BBP})}(t) $ to the ``generator'' operator
$G_{({BBP})}(t)$ which contains, due to the time-dependence of the
Dyson's map, also a Coriolis-force correction
${\Sigma}_{({BBP})}(t)$.

The second important warning concerns an innocent-looking but
deceptive subtlety as discussed more thoroughly in
Ref.~\cite{SIGMAdva}. Its essence is that the apparently independent
F-space ketket solutons of the apparently independent
Eq.~(\ref{SEtaujobe}) are just the S-space physical conjugates of
the usual F-space kets of Eq.~(\ref{SEtaujoa}). This means that
whenever one works in ${\cal H}^{(F)}$, one has to evaluate the
expectation values of a generic, hiddenly Hermitian observable
$A_{({BBP})}(t)$ using the F-space formula
 \be
 \bbr \Psi^{({BBP})}(t)|A_{({BBP})}(t)|\Psi^{({BBP})}(t)\kt\,
 \label{meva}
 \ee
where F-kets $|{\Psi}^{({BBP})}(t)\kt$ and
$|{\Psi}^{({BBP})}(t)\kkt= \Theta_{(BBP)}(t)|{\Psi}^{({BBP})}(t)\kt$
represent just an S-ket and its Hermitian S-conjugate, i.e., just
{\em the same} physical quantum state.


\section{Evolution in Heisenberg picture\label{secondly}}

In a Gedankenexperiment  one may prepare the Universe, at some
post-Big-Bang time $T>0$, in a pure state represented by a
biorthogonal pair of Hilbert-space elements
$|{\Psi}^{({BBP})}(T)\kt$ and $|{\Psi}^{({BBP})}(T)\kkt$. In such a
setting we may let the time to run backwards. Then we may solve
Eqs.~(\ref{SEtaujoa}) and (\ref{SEtaujobe}), in principle at least.
This might enable us to reconstruct the past, i.e., we could specify
the states of our Universe $|{\Psi}^{({BBP})}(t)\kt$ and
$|{\Psi}^{({BBP})}(t)\kkt$ at any $t>\tau_{(EP)}=0$.

\subsection{Heisenberg equations  }
\label{secs4s1}

The consistent picture of the unfolding of the Universe after Big
Bang cannot remain restricted to the description of the evolution of
wave functions. The test of the predictive power of the theory can
only be provided via a measurement, say, of the probabilistic
distribution of data. Thus, the theoretical predictions are
specified by the overlaps (\ref{meva}). By construction, the
variations of wave functions as controlled by the generator
$G_{({BBP})}(t)$ will interfere with the variations of the operator
$A_{({BBP})}(t)$ itself.

In our cosmological considerations the ``background of
quantization'' \cite{thie} characterizing the observable geometry of
the empty Universe is represented by the ``Alice-Bob distance''
operator $Q(t)$ or, in general, by a set of such operators. They are
assumed to be given as kinematical input, determining also the
time-dependent Dyson's map via Eq.~(\ref{thisco}). For all of the
other, dynamical observables  in ${\cal H}^{(F,S)}$, with formal
definition
 \be
 A_{({BBP})}(t)
 =\Omega_{({BBP})}^{-1}(t)
 \mathfrak{a}_{(SP)}(t){\Omega}_{({BBP})}(t)
 \label{heieq}
 \ee
a new problem emerges whenever they happen to be specified just at
an ``initial''/``final'' time $t=T$ of the preparation/filtration of
the quantum state in question. Still, the reconstruction of mean
values (\ref{meva}) remains friendly and feasible in Heisenberg
representation in which the wave functions are constant so that we
must set $G(t)=0$ and $H(t)=\Sigma(t)$ (cf. Ref.~\cite{Heisenberg}
for more details).

Naturally, whenever we decide to turn attention to the more general
non-adiabatic options with $G(t) \neq 0$, the above most convenient
assumption of our input knowledge of the map ${\Omega}_{({BBP})}(t)$
may prove too strong. With the purpose of weakening it we may
rewrite Eq.~(\ref{diffe}) in the Cauchy-problem form
 \be
 {\rm i}\partial_t\Omega_{({BBP})}(t)
 = \Omega^{}_{({BBP})}(t)
  {\Sigma}_{({BBP})}(t)
  \,
 \label{udiffe}
  \ee
to be read as a differential-equation definition of mapping
${\Omega}_{({BBP})}(t)$ from its suitable initial value (say, at
$t=T$) and from the more natural input knowledge of the Coriolis
force ${\Sigma}_{({BBP})}(t)$ of Eq.~(\ref{diffe}) which strongly
resembles (possibly, perturbed) Hamiltonian in Heisenberg picture.

After a return to the Heisenberg-picture assumption $H(t)=\Sigma(t)$
let us now differentiate Eq.~(\ref{heieq}) with respect to time.
Once we abbreviate $ \partial_t
\mathfrak{a}_{(SP)}(t)=\mathfrak{b}(t)$ and define
 \be
  B_{({BBP})}(t)=\Omega_{({BBP})}^{-1}(t)\mathfrak{b}(t){\Omega}_{({BBP})}(t)\,,
  \ \ \ \ \ H_{({BBP})}(t)=\Sigma_{({BBP})}(t)
 \label{heieqbe}
 \ee
this yields the first rule {\it alias} Heisenberg evolution equation
 \be
 {\rm i}\partial_t A_{({BBP})}(t) = A_{({BBP})}(t) {H}_{({BBP})}(t)
 -{H}_{({BBP})}(t) A_{({BBP})}(t)
 +{\rm i}B_{({BBP})}(t)\,
 \label{deveta}
 \ee
and an accompanying, adjoint rule
 \be
 {\rm i}\partial_t A^\dagger_{({BBP})}(t)
  = A_{({BBP})}^\dagger(t) {H}_{({BBP})}^\dagger(t)
 -{H}_{({BBP})}^\dagger(t) A_{({BBP})}^\dagger(t)
 +{\rm i}B_{({BBP})}^\dagger(t)\,.
 \label{devetbe}
 \ee
Formally, both of them resemble the Heisenberg commutation relations
and contain an independent-input operator (\ref{heieqbe}).
Naturally, the latter operator might have been given an explicit
form of an $T \to F$ transfer of the anomalous time-variability of
our observable {\it whenever considered time-dependent\,} already in
Schr\"{o}diger picture. Nevertheless, once we follow the classics
\cite{Geyer} and once we treat any return $F \to T$ as prohibited
(otherwise, the Dyson's non-unitary mapping would lose its {\it
raison d'\^{e}tre}), ``definition'' (\ref{heieqbe}) is inaccessible.
Due to the kinematical origin of Eqs.~(\ref{deveta}) or
(\ref{devetbe}), our knowledge of operator $B_{({BBP})}(t)$ at all
times must really be perceived as an {\it independent source of
input information} about the dynamics.

The list of the evolution equations for a quantum system in question
becomes completed. Naturally, the initial values of operators
${\Theta}_{({BBP})}(T)$ and $A_{({BBP})}(T)$ must be such that \be
 A^\dagger_{({BBP})}(T){\Theta}_{({BBP})}(T)
 ={\Theta}_{({BBP})}(T)A_{({BBP})}(T)\,
 \label{nejas}
 \ee
We may conclude that whenever $G(t)=0$, the construction of any
concrete toy model only requires the solution of Heisenberg
evolution Eqs.~(\ref{deveta}) or (\ref{devetbe}).


\subsection{The limitations of the Heisenberg picture of the Universe }
\label{secs4s2}

Before recalling any examples let us re-emphasize that the
Heisenberg representation {\it alias} Heisenberg picture (HP) of the
quantum systems provides one of the most straightforward forms of
hypothetical transitions between classical and quantum worlds. One
should immediately add that the HP approach proves extremely tedious
in the vast majority of practical calculations. It replaces the
dynamics described by the SP Schr\"{o}dinger equation for
wavefunctions by its much more complicated operator,
Heisenberg-equation equivalent. At the same time, once we are given
our ``geometry'' observable $Q(t)$ in its time-dependent
Heisenberg-representation form {\em in advance} (say, in a way
motivated, somehow, by the principle of correspondence), our tasks
get perceivably simplified.

In the underlying theory one assumes, therefore, that the set of the
admissible (and measurable) instantaneous quantized distances
$q(t)=q_n(t)$ between the two observers of Fig.~\ref{kruhym} are
eigenvalues of an operator $Q=Q(t)$ in some physical Hilbert space
${\cal H}^{(S)}$. This space is assumed endowed with the
instantaneous physical inner product which is determined, say, by a
time-independent metric $\Theta\neq \Theta(t)$ \cite{Heisenberg}. In
the case of a pure-state evolution, the integer subscript
$n=1,2,\ldots,N$ with $N\leq \infty$ may be kept fixed via a
preparation or measurement over the system at a time $t=T$.



Our quantum description of the Universe shortly after Big Bang will
be based, as already indicated above, on a non-Dirac, BBP amendment
of the Hilbert-space metric, on its factorization (\ref{fakto}) and
on the use of preconditioning of the ``clumsy'' physical wave
function $|\psi(t)\pkt \in {\cal H}^{(T)}$ of the Universe,
 \be
 |\psi(t)\pkt=\Omega_{(BBP)}(t)\,|\psi(t)\kt
 =\left [\Omega_{(BBP)}^\dagger(t)\right ]^{-1}|\psi^{}(t)\kkt
 \,.
  \label{BBouSE}
 \ee
(cf. Eq.~(\ref{precon}) below, and note also the unfortunate typo in
equation Nr.~(7) of Ref.~\cite{Heisenberg} where the exponent
$(^{-1})$ is missing).

As long as the mapping $\Omega$ is allowed time-dependent, the
standard Schr\"{o}dinger equation which determined the evolution of
a pure state $|\psi(t)\pkt $  in space ${\cal H}^{(T)}$ in
Schr\"{o}dinger picture cannot be replaced by Eq.~(\ref{SEtd})
anymore. Indeed, one must leave the standard Schr\"{o}dinger picture
as well as its non-Hermitian stationary amendment and
implementations as described in Refs.~\cite{Geyer,Carl,ali}.

Secondly, without additional assumptions one cannot employ the
non-Hermitian Heisenberg picture, either. The reason is that in this
framework (in which the observables are allowed to vary with time)
the Hilbert space metric must still be kept constant
\cite{Heisenberg}. Thus, our theoretical quantum description of the
evolution of the Universe in Heisenberg picture must be accompanied
by the adiabaticity assumption $\partial_t\Theta(t) =$ small.

\section{What could have happened before Big Bang?\label{thirdly}}

The applicability of the above-summarized crypto-Hermitian version
of Heisenberg picture of Ref.~\cite{Heisenberg} may be now sampled
by any above-mentioned schematic toy model of the Universe in
adiabatic approximation. Operator $Q(t)$ (defined as acting in a
preselected Hilbert space ${\cal H}^{(F)}$) is assumed given (or
guessed, say, on the background of correspondence principle) in
advance, as a tentative input information about dynamics.

In addition, our schematic Universe living near Big Bang may be also
endowed with an additional pair of observables $A$ and $B$, with
their mutual relation clarified by the pair of Eqs.~(\ref{heieq})
and (\ref{heieqbe}). In principle, in the light of
Eq.~(\ref{deveta}) the  former operator may be specified just at the
initial time $t=T$. In this sense the models with the necessity of
specification of $B(t) \neq 0$ {\em at all times} may be considered
anomalous (cf. also the related discussion in \cite{Heisenberg}).

Naturally, even if we assume that $B(t) = 0$, the solution of
Heisenberg Eq.~(\ref{deveta}) need not be easy. For this reason, we
shall now display the results of a quantitative analysis of a few
most elementary models. We shall employ the following simplifying
assumptions: (1) In the spirit of Fig.~\ref{kruhym}, only the
quantized distance between Alice and Bob (i.e., just a single
geometry-representing and adiabatically variable observable $Q(t)$)
will be considered. (2) For the sake of simplicity, our illustrative
samples of the kinematical input information (i.e., of the operators
$Q(t)$) will only be considered in a finite-dimensional, $N$ by $N$
matrix form, $Q(t)=Q^{(N)}(t)$.

\begin{figure}                    
\begin{center}                         
\epsfig{file=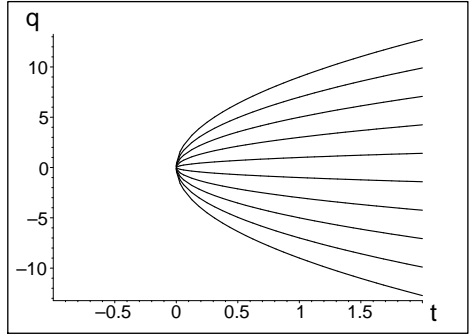,angle=270,width=0.35\textwidth}
\end{center}   
\vspace{2mm} \caption{Real eigenvalues of the toy-model geometry
(\ref{hoyma}) in the both-sided vicinity of its Big Bang singularity
at $t=0$.
 \label{obyc}
 }
\end{figure}

\subsection{No tunneling and no observable space before Big Bang}
\label{secs5s1}

For illustration purposes let us first recall the $N$ by $N$ real
matrix model of Refs.~\cite{maximal} with
 \be
 Q^{(N)}(t)=Q_0^{(N)}+\sqrt{1-t}\times Q_1^{(N)}
 \label{hoyma}
 \ee
which is composed of a diagonal matrix $Q_0^{(N)}$ with equidistant
elements
 \be
 \left [Q_0^{(N)}\right ]_{nn} = \{-N+1, -N+3, \ldots, N-1 \}
 \ee
and of an antisymmetric time-dependent ``perturbation'' with a
tridiagonal-matrix coefficient  $Q_1^{(N)}$ with zero diagonal and
non-vanishing elements $\left [Q_0^{(N)}\right ]_{n+1,n}
 =- \left [Q_0^{(N)}\right ]_{n,n+1} =$
 \be
  =\{\sqrt{1\cdot (N-1)}, \sqrt{2\cdot (N-2)},\sqrt{3\cdot (N-3)},
 \ldots,\sqrt{(N-1)\cdot 1},  \}\,.
 \ee
In Refs.~\cite{maximal} the choice of this model was dictated by its
property of having real and equidistant spectrum at all of the
non-negative times $t>0$. Another remarkable feature of this model
is that while matrix (\ref{hoyma}) is real and manifestly
non-Hermitian at all times $t\in (-\infty,1)$, it becomes diagonal
at $t=1$ and complex and Hermitian at all the remaining times $t\in
(1,\infty)$.

At $N=10$ the spectrum of such a toy model is sampled in
Fig.~\ref{obyc}. Obviously, this example of a  kinematical input
connects, smoothly, the complete Big-Bang-type degeneracy of the
eigenvalues at $t=0$ with their unfolding at $t>0$ which passes also
through the ``unperturbed'', diagonal-matrix special case at $t=1$.
Needless to emphasize that in this model the spectrum is all complex
and, hence, the space of the Universe remains completely
unobservable {\it alias} non-existent before Big Bang.

\subsection{Cyclic cosmology}
\label{secs5s2}

\begin{figure}[h]                    
\begin{center}                         
\epsfig{file=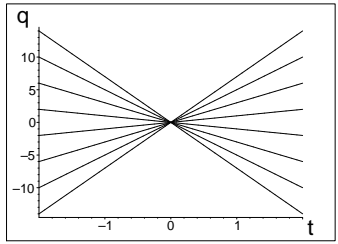,angle=270,width=0.25\textwidth}
\end{center}     
\vspace{2mm} \caption{Real eigenvalues of the toy-model geometry
(\ref{royma}) in the both-sided vicinity of its Big Bang singularity
at $t=0$.
 \label{cycl8}
 }
\end{figure}

 \noindent
Not quite expectedly the spectrum gets entirely different after an
apparently minor change of the time-dependence in
 \be
 Q^{(N)}(t)=Q_0^{(N)}+\sqrt{1-t^2}\times Q_1^{(N)}
 \label{royma}
 \ee
Using $N=8$ the resulting spectrum is displayed in Fig.~\ref{cycl8}.
We see that in the new model the ``geometry of the world'' was the
same before Big Bang so that model (\ref{royma}) may be perceived as
reflecting a kinematics of a kind of cyclic cosmology as preferred
in Hinduism or, more recently, by Roger Penrose \cite{Penrose}.

\subsection{Darwinistic, evolutionary cosmology}
\label{secs15}

In the THS representation of the 1D Universe the ``geometry'' or
``kinematical'' operator $Q(t)$ may be assumed, in general,

\begin{itemize}

\item
non-Hermitian (otherwise, we would lose the dynamical degrees of
freedom carried by the generic metric $\Theta$ and needed and
essential near the Big Bang instant),

\item
simple (i.e., typically, tridiagonal as above -- otherwise, there
would be hardly any point in our leaving the much simpler
Schr\"{o}dinger picture).

\end{itemize}

 \noindent
In the latter sense, our third class of toy models may be taken from
Refs.~\cite{Wu} and ~\cite{Borisov} and sampled by the following
$N=8$ distance operator
 \be
 Q(t)=\left[ \begin {array}{cccccccc} 0&1-t&0&0&0&0&0&0\\\noalign{\medskip}
t&0&1-t&0&0&0&0&0\\\noalign{\medskip}0&t&0&1- \left| t \right|
&0&0&0 &0\\\noalign{\medskip}0&0& \left| t \right| &0&1- \left| t
\right| &0&0 &0\\\noalign{\medskip}0&0&0& \left| t \right| &0&1-
\left| t \right| &0 &0\\\noalign{\medskip}0&0&0&0& \left| t \right|
&0&1-t&0
\\\noalign{\medskip}0&0&0&0&0&t&0&1-t\\\noalign{\medskip}0&0&0&0&0&0&
t&0\end {array} \right]\,
 \label{toyma}
 \ee
The piecewise linear time-dependence of this operator leads to the
quantum phase transition between the Big Crunch collapse of the
spatial grid in previous Eon and the Big Bang start of the spatial
expansion of the present Eon. In the vicinity of the singularity at
$t=0$ we may characterize such a quantum cosmological toy model by
the following flowchart,
 \ben
 \\
 \ba
    \ \ \
    \ \ \ \ \
    \ \ \ \ \
       \begin{array}{|c|}
 \hline
   {\rm {\bf  working}\  space\ }  {\cal H}^{(F)} \\
   {\rm  the\ observable\ of\ {\bf  geometry}\ }  Q(t) \\
     {\rm  defined \ at\ all\ real\ times } \\
     {\rm   {\bf  non-diagonalizable} \ at\  } t_{(EP)}=0\\
 \hline
 \ea
 \ \ \ \ \
 \ \ \ \ \
 \ \ \ \ \
       \\
 \\
 \stackrel{{\bf previous\ Eon, } }{t<0}\ \swarrow
 \ \ \ \
 \ \ \ \
 \ \ \ \
 \ \ \ \
 \ \  \ \searrow \ \ \
 \stackrel{{\bf our \ Eon, } }{t>0}
 \ \ \ \
   \ \ \\
 \\
  \ \ \  \
   \ \ \ \
 \begin{array}{|c|}
 \hline
 {\rm  {\bf underdeveloped\ }standard\ space\ }
    {\cal H}^{({S'})}_{} \\
  {\rm   {\bf  ghosts\ }to\ be\  projected\ out} \\
  {\rm (some\ eigenvalues\ {\bf not\ yet}\ observable)} \\
  \hline
 \ea
  \ \ \  \  \  \ \ \ \
 \begin{array}{|c|}
 \hline
  {\rm   {\bf our \ } Hilbert\ space} \
    {\cal H}^{(S)}_{} \\
   {\rm observable\ } {Q}= {Q}^\ddagger =\Theta_S^{-1}Q^\dagger \Theta_S\\
 {\rm (all\  eigenvalues \ {\bf real})} \\
 \hline
 \ea
 \ \ \ \ \
 \ \ \ \ \
 \ \ \ \ \
  \\
 \\
 \ \swarrow \!\!\! 
  \ \  \ \ \ \
 \stackrel{{\bf auxiliary\ maps\  to \ Schroedinger \ picture} }{} \
  \ \ \ \searrow \!\!\! 
   \ \ \
   \ \ \\
   \\
 \begin{array}{|c|}
 \hline
   {\rm   {  }third\  space} \
    {\cal H}^{(T')} \\
 {\rm  of\ the\  {\bf extinct\ } Universe} \\
 \hline
 \ea
 \
      \ \ \ \ \ \ \ \ \ \ \ \ \ \ \ \ \ \ \ \ \ \ \
            \
       \begin{array}{|c|}
 \hline
  {\rm   {} third\ space} \
    {\cal H}^{(T)} \\
 {\rm   {\bf contemporary\ }Universe  } \\
 \hline
 \ea
 \,.
  \ \ \ \ \ \
 \\
   \ea
   \\
   \\
 \een
The evolutionary-cosmology idea of the quantum Crunch-Bang
transition itself (discussed more thoroughly in Ref.~\cite{Borisov}
and illustrated also by Fig.~\ref{sy}) may be perceived as one of
the serendipitous conceptual innovations provided by the present
Heisenberg-picture background-independent \cite{thie} quantization
of our schematic Universes.

\begin{figure}                    
\begin{center}                         
\epsfig{file=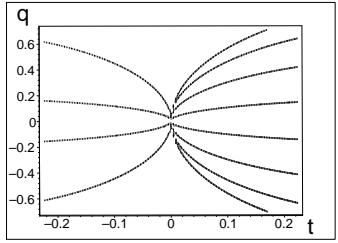,angle=270,width=0.25\textwidth}
\end{center}   
\vspace{2mm} \caption{Real eigenvalues of the toy-model geometry
(\ref{toyma}) in the both-sided vicinity of the Crunch-Bang
singularity at $t=0$.
 \label{sy}
 }
\end{figure}

\section{Outlook \label{fourthly}}

The results of the analysis of the solvable models of preceding
section offer a nice illustration of several merits of the THS
approach to the building of Big-Bang-exhibiting quantum systems.

\begin{itemize}

\item
the Big Bang value of time $t^{(BB)}=0$ is a point of degeneracy of
all of the eigenvalues, $q_n(0)=0$ at all $n=1,2,\ldots,N$;

\item
at $t=t^{(BB)}=0$ all of our toy models acquire the complete, $N$ by
$N$ Jordan-block structure so that the Big Bang time coincides with
the point of confluence of {\em all} of the Kato's  exceptional
points;

\item
after Big Bang, i.e., at  $t>t^{(BB)}=0$ the spectra of possible
(and growing) quantum distances between Alice and Bob are all real
and, hence, observable, in our specific toy models at least;

\item
in the light of Fig.~\ref{sy} our models describe also the times
{\em before} Big Bang, $t<t^{(BB)}=0$. In this sense the pass of our
systems through the Big-Bang singularity is ``causal'', described by
a ``universal'' operator $Q(t)$;

\item
before Big Bang (i.e., before the Big Crunch of the Penrose's
\cite{Penrose} ``previous Eon'') the menu of the {\em real}
distances $q_n(t)$ is replaced by an empty set (in Fig.~\ref{obyc}),
survives unchanged (in Fig.~\ref{cycl8}) or gets reduced to a proper
subset (cf. Fig.~\ref{sy});

\item
in the most interesting latter case the ``missing'', complex
eigenvalues are tractable as ``not yet observable''. One could speak
about various ``evolutionary'' forms of cosmology in this setting.

\end{itemize}


\section*{Appendix. Auxiliary spaces and ${\cal PT}$ symmetries}
\label{secsa1}

A few years after the publication of review~\cite{Geyer}, a series
of rediscoveries and an enormous growth of popularity of the pattern
followed the publication of pioneering letter \cite{BB} in which
Bender with his student inverted the flowchart. They choose a nice
illustrative example to show that the {\em manifestly non-Hermitian}
$F-$space Hamiltonian $H$ with real spectrum may be interpreted as a
hypothetical {\em input} information about the dynamics (cf. also
review \cite{Carl} for more details).

Graphically, the flowchart of ${\cal PT}-$symmetric quantum theory
is schematically depicted in Fig.~\ref{ptpo4}. For completeness let
us add that the Bender's and Boettcher's construction was based  on
the assumption of ${\cal PT}$ symmetry $H {\cal PT} = {\cal PT}H$ of
their dynamical-input Hamiltonians where the most common
phenomenological parity ${\cal P}$ and time reversal ${\cal T}$
entered the game. Mostafazadeh (cf. his review \cite{ali})
emphasized that their theory may be generalized while working with
more general ${\cal T}$s (typically, any antilinear operator) and
${\cal P}$s (basically, any indefinite, invertible operator).


Several mathematical amendments of the theory were developed in the
related literature, with the main purpose of making the
constructions feasible. Let us only mention here that the useful
heuristic role of operator ${\cal P}$ was successfully transferred
to the Krein-space metrics $\eta$ (cf. \cite{AKbook} for a
comprehensive review). In comment \cite{shendr} we explained that in
principle, the role of ${\cal P}$ could even be transferred to some
positive-definite, simplified and redundant auxiliary-Hilbert-space
metrics $\tilde{\cal P} = \Theta_A \neq \Theta_S$. Such a recipe
proved encouragingly efficient \cite{diagmetr}. Its flowchart may be
summarized in the following diagram
 \ben
 \\
 \ba
    \ \ \
       \begin{array}{|c|}
 \hline
  \fbox{\rm input:}\\
  {\bf   space} \
    {\cal H}^{(F)} \ {\rm is}\  {\bf friendly  }\\
 {\bf  metric\ } \Theta^{(Dirac)}=I\ {\rm is}\   {\bf false }
 \\
  {\rm   observable\ }
   {{Q}(t)}\neq  {{Q}}^\dagger(t)\ {\rm is} \ given\\
 \hline
 \ea
 \ \ \ \ \
 \ \ \ \ \
 \ \ \ \ \
       \\
 \\
 \ \ \ \
 \stackrel{{\bf  \ preliminary\ Dyson\  map\ } }
 \ \swarrow
 \ \ \ \ \ \
 \ \ \ \ \ \
 \ \ \ \
 \ \ \ \
 \ \  \ \searrow \ \ \
 \stackrel{{\bf  \  correct\ Dyson\  map\ } }
 \ \ \ \
 \ \ \ \ \ \
 \ \ \ \ \ \
   \ \ \\
 \\
  \ \ \  \
 \begin{array}{|c|}
 \hline
  \fbox{\rm  reality\ proof:} \\
 {\rm  {\bf artificial\ }space\ }
    {\cal H}^{(A)}_{} \\
     {\rm  {\bf auxiliary \ } }
    \Theta_A=\Omega_A^\dagger \Omega_A
        \\
             {Q}^\sharp =\Theta_A^{-1}Q^\dagger \Theta_A
             ={Q}
    \\
  \hline
 \ea
  \ \ \  \  \  \ \
 \ \stackrel{{\bf   not\ related\ to } }{}\
   \ \ \ \ \ \
 \begin{array}{|c|}
 \hline
  \fbox{\rm output:} \\
  {\rm   {\bf standard \ } space} \
    {\cal H}^{(S)}_{} \\
     {\bf correct\  } \Theta_S=\Omega_S^\dagger \Omega_S
    \\
    {Q}^\ddagger =\Theta_S^{-1}Q^\dagger \Theta_S={Q}\\
 \hline
 \ea
 \ \ \ \ \
 \ \ \ \ \
 \ \ \ \ \
  \\
 \\
 \ \swarrow \!\!\! \nearrow \
  \ \  \ \ \ \
  \ \  \ \ \ \
  \ \  \ \ \ \
 \stackrel{{\bf unitary\  equivalences} }{} \
  \ \  \ \ \ \
  \ \  \ \ \ \
  \ \ \ \searrow \!\!\! \nwarrow \ \ \ \
   \ \ \\
   \\
 \ \begin{array}{|c|}
 \hline
   {\rm   {\bf  byproduct\ } } \
    {\cal H}_{(math.)} \\
      {\mathfrak{q}}_{(math.)}=
      \Omega_A
      Q
      \Omega_A^{-1}
      = {\mathfrak{q}}_{(math.)}^\dagger\\
 {\rm   (redundant)} \\
 \hline
 \ea
 \ \
      \ \stackrel{{\bf   not\ related\ to } }{}\
            \
       \begin{array}{|c|}
 \hline
  {\rm   {\bf textbook\ } space} \
    {\cal H}^{(T)}_{(phys.)} \\
     {\mathfrak{q}}_{(phys.)}= {\mathfrak{q}}^\dagger_{(phys.)} =
   {\rm \ realistic\ }
     \\
 {\rm (inaccessible)  }\\
 \hline
 \ea
 \ \ \ \ \ \ \ \
 \\
   \ea
   \\
   \\
 \label{FHS}
 \een
Besides the right-side flow of mapping we see here the auxiliary,
unphysical left-side flow where, typically, the non-Dirac metric
$\Theta_A$ need not carry any physical contents. In some models such
an auxiliary metric proved even obtainable in a trivial
diagonal-matrix form \cite{Borisov}.

\begin{figure}
\begin{center}                         
\epsfig{file=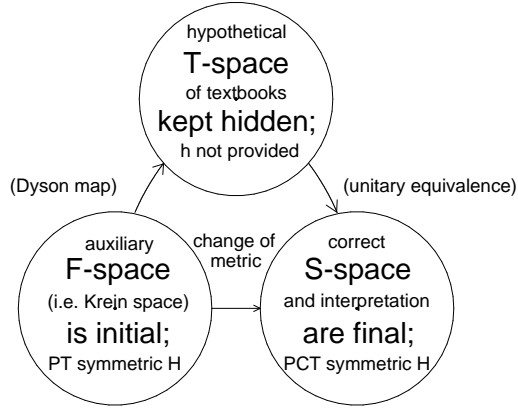,angle=270,width=0.55\textwidth}
\end{center}      
\vspace{2mm} \caption{THS interpretation of ${\cal PT}-$ symmetric
Hamiltonians $H$.
 \label{ptpo4}
 }
\end{figure}

\end{document}